\documentstyle[twocolumn,aps,floats]{revtex}

\begin{document}

\draft \tolerance = 10000

\setcounter{topnumber}{1}
\renewcommand{\topfraction}{0.9}
\renewcommand{\textfraction}{0.1}
\renewcommand{\floatpagefraction}{0.9}

\twocolumn[\hsize\textwidth\columnwidth\hsize\csname
@twocolumnfalse\endcsname

\title{ Do  Gravitational and Electromagnetic
 Fields Have  Rest Masses in the Fractal Universe? }
\author{L.Ya.Kobelev  \\ Department of  Physics, Urals State University \\
 Av. Lenina, 51, Ekaterinburg 620083, Russia \\  E-mail: leonid.kobelev@usu.ru}
\maketitle

\begin{abstract}
As well known  rest masses of  elementary particles and physical fields
appear when temperature of Universe become low enough and symmetry broke.
Are there another sources of rest masses that are consequences of other
nature of rest masses or it is the only methods for generating rest
masses? What will be with massless fields when the laws of symmetry are
not exact laws but only are very good approximation and the time is not
homogeneous, as it is in the fractal world? In this paper  based on the
fractal theory of time and space (developed by author earlier)  a possible
source of rest masses caused by the fractional dimensions (FD) of the time
is considered. It gives rest masses (very small) for all particles and
fields including gravitational and electromagnetic fields. The estimation
of the values of the rest masses  gives  $m_{g}=\sqrt{\varepsilon
(tt_{i})^{-1}}$, $m_{f}= \sqrt{(t_{0}t_{i})^{-1}}$ where  $t_{i}$ is the
necessary time for photons and gravitons with rest masses $m_{ph}$ and
$m_{g}$  to  get the velocity equal the speed of light (in the fractal
Universe the moving with any velocities is possible), $t_{0}$ is the time
of existence of Universe. Under some assumption preliminary values of rest
masses photons and gravitons  are obtained:  for a rest mass of photons $
m_{f}\sim 10^{-39}g$, for a rest mass of gravitons $m_{g}\sim10^{-43}g$.

\end{abstract}

 \vspace{1cm}

]

\section{Introduction}

The problems of existence rest masses  of electromagnetic, gravitational
and neutrino fields are open till now. From one side the experiments can
not discover rest masses of these fields, from other side there is
unexplained question why some physical fields have rest masses, the other
fields  not have it.In early years it was wide-spread the opinion that the
fields of the interactions (electromagnetic, gravitational an so on)
between particles are massless and must not take a charges, but example of
the electro-weak interactions by means of  vector mesons broke this
opinion and interactions of quarks broke the opinions about chargeless of
interactions fields. Thus the situation is: there are the physical fields
with rest masses and there are the physical fields without rest masses;
there are the interactions fields  with charges and without charges. There
are the theorems  (Goldstone and analogies) about massless boson fields as
consequence of symmetry laws. In the space with the fractal time (see
\cite{kob1}- \cite{kob15}  not all the symmetry laws are fulfilled and
these theorems  work only as a good approach. Only in the case when the
fractal dimension of time almost integer the symmetry laws are almost
fulfill, not exactly, but as a good approach. So the existence in the
fractal world rest masses of bosons field { such as electromagnetic and
gravitational) do not contradicts any theorems true for world with exact
symmetry laws and homogeneous time. In this paper shown that in the space
with fractal time (dimensions of time $d_{t}= 1+ \varepsilon ( {\bf
r},t)\neq 1)$ in the frame-work of the fractal theory of time and space
\cite{kob1}- \cite{kob15} all physical fields have rest masses originated
by fractional additives $\varepsilon$ to integer time dimension. For the
fields with rest masses in the integer dimensions time these additional
masses are non-essential, but it gives the main masses for the  massless
early fields such as electromagnetic and gravitational fields. So the
symmetry for all physical fields in respect to rest masses is restored: in
fractal Universe all physical fields have rest masses: some fields have
rest masses originated  when temperature of Universe become low enough and
symmetry laws is broken, other fields got rest masses by means of
fractional dimensions of time (or space) from the time of beginning of the
"big bang", when the time and space with fractal dimensions were
originated.

\section{  How  to describe the functions with dependencies at time and
coordinates in the fractal world? }

As the first step of describing the physical functions in the fractal
Universe (the world with fractional dimensions of time and space) it is
necessary to introduce the definitions of fractal time and fractal space
(for the purpose of this paper it is enough to introduce only the fractal
time) and mathematical methods its describing. Following \cite{kob1},
\cite{kob2}, we will consider both time and space as the initial real
material fields existing in the world and generating all other physical
fields. Therefore, we  introduce the integral functionals (both left-sided
and right-sided) which are suitable to describe the dynamics of functions
defined on multifractal sets of time and space (generalized fractional
derivatives (GFD), see \cite{kob1}-\cite{kob2}, \cite{kob3}) and replace
by GFD the usual derivatives and integral respect to time and space
coordinates  in the fractional dimensions functional. These functionals
GFD are simple and natural generalization of the Riemann-Liouville
fractional derivatives and integrals:
\begin{equation} \label{1}
D_{+,t}^{d}f(t)=\left( \frac{d}{dt}\right)^{n}\int_{a}^{t}
\frac{f(t^{\prime})dt^{\prime}}{\Gamma
(n-d(t^{\prime}))(t-t^{\prime})^{d(t^{\prime})-n+1}}
\end{equation}
\begin{equation} \label{2}
D_{-,t}^{d}f(t)=(-1)^{n}\left( \frac{d}{dt}\right)
^{n}\int_{t}^{b}\frac{f(t^{\prime})dt^{\prime}}{\Gamma
(n-d(t^{\prime}))(t^{\prime}-t)^{d(t^{\prime})-n+1}}
\end{equation}
where $\Gamma(x)$ is Euler's gamma function, and $a$ and $b$ are some
constants from $[0,\infty)$. In these definitions, as usually, $n=\{d\}+1$
,where $\{d\}$ is the integer part of $d$ if $d\geq 0$ (i.e. $n-1\le d<n$)
and $n=0$ for $d<0$. If $d=const$, the generalized fractional derivatives
(GFD) (\ref{1})-(\ref{2}) coincide with the Riemann - Liouville fractional
derivatives ($d\geq 0$) or fractional integrals ($d<0$). When
$d=n+\varepsilon (t), \varepsilon (t)\rightarrow 0$, GFD can be
represented by means of integer derivatives and integrals.We pay attention
that there are relations between GFD and ordinary derivatives for
$d_{\alpha}$ near integer values. If $d_{\alpha}$ $\rightarrow n$ where
$n$ is an integer number, for example $d_{\alpha}$ = $1+\varepsilon({\b r}
(t),t)$, $ \alpha={\b r} ,t $, in that case it is possible represent GFD
by approximate relations (see \cite{kob1}- \cite{kob3})
\begin{equation}\label{3}
  D_{+,x_{\alpha}}^{1+\epsilon}f({\b r} (t),t)=
  \frac{\partial}{\partial{x_{\alpha}}} f({\b r} (t),t)+
  \frac{\partial}{\partial x_{\alpha}}[{\varepsilon
  ({\b r}(t),t)f({\b r}(t),t)]}
\end{equation}
For $n=1$, that is, $d=1+\varepsilon$, $\left| \varepsilon \right| <<1$ it
is possible to obtain:
\begin{equation} \label{4}
D_{+,t}^{1+\varepsilon }f(t)\approx \frac{\partial}{\partial t}
f(t)+a\frac{\partial}{\partial t}\left[\varepsilon (r(t),t)f(t)\right]
\end{equation}
where $a$ is constant and defined by the choice of the rules of
regularization of integrals (\ref{1})-(\ref{2}) (for more detailed see
\cite{kob1},\cite{kob2}, \cite {kob3}). The selection of the rule of
regularization that gives a real additives for usual derivative in
(\ref{3}) yield $a=0.5$ for $d<1$ \cite{kob1}. The functions under
integral sign in (\ref{1})-(\ref{2}) we consider as the generalized
functions defined on the set of the finite functions \cite{gel}. The
notions of GFD, similar to (\ref{1})-(\ref{2}), can also be defined and
for the space variables ${\mathbf r}$. In the definitions of GFD
(\ref{1})-(\ref{2}) the connections between fractal dimensions of time
$d_{t}({\mathbf r}(t),t)$  and main characteristics of physical fields
(say, potentials $\Phi _{i}({\mathbf r}(t),t),\,i=1,2,..)$ or densities of
Lagrangians $L_{i}$) are determined. Following \cite{kob1}, we define this
connection by the relation
\begin{equation} \label{5}
d_{t}({\mathbf r}(t),t)=1+\sum_{i}\beta_{i}L_{i}(\Phi_{i} ({\mathbf
r}(t),t))
\end{equation}
where $L_{i}$ are densities of energy of physical fields, $\beta_{i}$ are
dimensional constants with physical dimension of $[L_{i}]^{-1}$ (it is
worth to choose $\beta _{i}^{\prime}$ in the form $\beta _{i}^{\prime
}=a^{-1}\beta _{i}$ for the sake of independence from regularization
constant). The definition of time as the system of subsets and definition
the FD $d$ (see  \ref{4}) connects the value of fractional (fractal)
dimension $d_{t}(r(t),t)$  with each time instant $t$. The latter depends
both on time $t$ and coordinates ${\mathbf r}$. If $d_{t}=1$ (an absence
of physical fields) the set of time has topological dimension equal to
unity.\\ Thus, in the fractal Universe for changes of any physical
functions describing it is necessary to use the GFD instead of ordinary
derivatives and integrals. Only in the case when corrections to integer
dimensions is small (it is the wide-spread case in our world) it is
possible replace the GFD by ordinary derivatives using the relations
(\ref{3}). We  stress now that GFD $D_{+,t}^{d_{t}} $ take into account
the influences of the  past of the phenomena on the present time state of
the phenomena. The GFD $D_{-,t}^{d_{t}}$ take into account the influences
of future times on the present time state of the phenomena. So the
mathematical methods of GFD defined on the multifractal sets allow to take
into account the influences of past and future on the state of system at
the present time. In  the fractal world describing by GFD there are no
stable phenomena because the differentiation by means of GFD of constant
value do not gives zero.

\section{The fractal Universe is in continual changing and  expanding }

The hypotheses of the "big bang " as the originate of our  Universe gives
continual changing and expanding of time and space, but the mathematical
methods of ordinary differential and integral calculus  based on usual
mathematical analysis do not give possibility to describe this changing
correctly. In particular, thou the time and space are continually
expanding (in the frame of GR,) from point of view of ordinary analysis in
this expanding world exist constant physical values (for example a volume,
an energy and so on). The fractal Universe is an open system (about open
systems see \cite{klim1}, \cite{klim2}) defined on the measure carrier
(see \cite{kob15}, \cite{kob1}-\cite{kob15}). It is expanding, so all
physical values in our world must not be constant. Thou its the  changing
is very small it  has a principle role. The GFD gives such changing of all
physical phenomena and values. That changes originate by fractional
dimensions of time ( and the space) and allow to describe the changing
concerned with Universe expanding. Let us see now how to calculate the GRD
(\ref{1}) with respect to $t$ at constant value $a_{0}=const$. In the
fractal world there are no constant values because GFD of constant values
do not equal zero
\begin{equation}\label{6}
 D_{+,t}^{1+\varepsilon}a_{0}=\frac{a_{0}\varepsilon}{t^{1+\varepsilon}\Gamma(1+\varepsilon)}
\approx \frac{a_{0}\varepsilon}{t^{1+\varepsilon}}
\end{equation}
for large $t$. So if our world has the fractal dimensions any values in
this world including the constant values of our world may be expanded in
infinite power series in $t$. The GFD $D_{-,t}^{d_{t}}$ of constant value
$a_{0}$ ($d_{t}=1+\varepsilon({\bf r},t),|\varepsilon|<<1 $)is equal
\begin{equation}\label{7}
   D_{-,t}^{d_{t}}a_{0} \approx \frac{a_{0}\varepsilon}{B-t}
\end{equation}
In (\ref{7}) the $B$ -is the upper limit of integral in GFD. The time $t$
in (\ref{1}) - (\ref{2}) is the time of existence of Universe. It is
useful for symmetry of (\ref{6}) and (\ref{7}) to choose $B$ equal $2t$ .
In that case (though it has no principle sense) the influence of future on
present time is bounded by the time equal the time of existence of
Universe.

\section{How the fractal dimensions of time
originate rest masses in the massless field theories  ?}

The  corrections given by GFD in case of using it for differentiation of
constant value were considered  early . Let us now represent (\ref {7}) in
the form when its main characteristics conserves but the derivatives are
the ordinary. Thus an any constant value $a_{0}$ in fractal world may be
represented as the value with dependence of time $t$ ( or coordinates
${\bf r}$)
\begin{eqnarray}\label{8}
  a\approx a_{0}\exp\frac{\varepsilon}{t^{\varepsilon}}
\end{eqnarray}
In (\ref{8}) the factor $exp(\varepsilon t^{-\varepsilon})$ describes the
corrections given by the fractal dimensions of time. The ordinary
differentiation of (\ref{8}) with respect to $t$ coincide with result of
GFD of $a$ with respect to $t$. If $d_{t}=1$ the time dependence in
(\ref{8}) disappears ($\varepsilon =0$ ). The (\ref{8}) may be used for
calculation of the ordinary derivatives in the theory of the fractal time
on base of relation (\ref{4}) if $d_{t}=1+\varepsilon({\bf r},t)$ and
$|\varepsilon|<<1$. In the fractal time and space the equations of all
physical fields must be wrote by means of GFD. For case when FD of time
has small fractal corrections to unit these equations may be wrote by
means of ordinary derivatives but using (\ref{4}). Than in the  equations
with GFD appear the members proportional to fields that usually describe
the rest masses. These members are result of existence of fractal
dimensions of time in fractal Universe where the time treats as the real
field and the field of time originate all physical fields ( see
\cite{kob1} - \cite{kob4}) including the rest masses of massless fields.
As the examples of originating the rest masses by GFD the demonstration of
appealing rest masses for electromagnetic and gravitational fields will be
considered in next paragraph .

\section{ The rest masses of photons and  gravitons in the  Universe
with fractal time}

The equations of electromagnetic and gravitational  fields in the fractal
Universe  if take into account both the corrections giving by FD and
modifying SR ( see also \cite{kob1},\cite{kob2}, \cite{kob10},
\cite{kob14} ,\cite{kob15}) reads: \\ \\a) electromagnetic fields
equations
\begin{eqnarray}\label{9}
[ D_{ - ,\bf {r}}^{d_{\bf r} } D_{ +{\bf r}}^{d_{\bf r} } - \frac{1} {{c^2
}}D_{ - ,t}^{d_t } D_{ + ,t}^{d_t }+ m^2 ]I A_\mu (x) = [
\alpha_{1}\frac{{4\pi }} {c}j_\mu(x)+\\ \nonumber+2a_{0}D_{ - ,t}^{d_t
}D_{ + ,t}^{d_t }\alpha_{2}A_{\mu}(i)] , \mu =0,1,2,3
 \end{eqnarray}
\begin{eqnarray}\label{A}  \nonumber
  {\bf j}_{i} = eD_{ +, t}^{d_{i} } {\bf r}_{i} ,\quad i=1,2,3
\end{eqnarray}
\begin{equation}\label{10}
D_{ + {\mu}}^{d_{\mu} } A_{\mu} (x) = 0
\end{equation}
b)   gravitational fields equations (the measure carrier is Minkowski
time- space)
\begin{eqnarray}\label{11}
  \gamma^{\alpha\beta}D_{-,\alpha}^{d_{i}}D_{+,\beta}^{d_{i}}I\tilde{\Phi}^{\mu
\nu} = \alpha_{1}b^{2}\tilde{\Phi}^{\mu\nu}+
  \lambda{\tilde t}^{\mu \nu }(\gamma^{\mu\nu},\Phi_{A})+\\ \nonumber
 + 2a_{0}\gamma^{44}D_{-,t}^{d_{t}}D_{+,t}^{d_{t}}\alpha_{2}\tilde{\Phi}^{\mu\nu}
\end{eqnarray}
In (\ref{10}) and (\ref{11})) $\alpha_{1}$, $\alpha_{2}$ are Dirac type
matrices, $I$ is unit four column matrix,  the function $a_{0}$ has the
form
\begin{eqnarray}\label{12}
  a_{0}= \varepsilon({\bf r}(t),t)= a_{g}+a_{e}+a_{n} &=& \nonumber \\
 = \beta_{g}L_{g}({\bf r},t)+\beta_{e}L_{e}({\bf
  r},t)+  \beta_{n}L_{n}({\bf r},t)
\end{eqnarray}
where $L_{g}$, $L_{e}$, $L_{n}$ are Lagrangians density of energies for
gravitational, electro-weak and strong fields, $\beta_{i}$ is the full
energies of the fields that constructed the  $L_{i}$ .\\
\\ We consider now the cases of free electromagnetic and gravitational
fields  and neglect by influences of the fields with imaginary energies
(this influences gives non-essentials corrections  see \cite{kob15}) and
omit the members with $m^{2}$ and $b^{2}$ ( which were formal introduced
for fulfillment of the conditions of existence the GFD ). Thus we have
\begin{equation}\label{14}
[ D_{ - ,\bf {r}}^{d_{\bf r} } D_{ +{\bf r}}^{d_{\bf r} } - \frac{1}
{c^{2}}D_{ -, t}^{d_{t} } D_{ + ,t}^{d_{t}} ]I A_\mu (x) =0
\end{equation}
\begin{eqnarray}\label{15}
  \gamma^{\alpha\beta}D_{-,\alpha}^{d_{i}}D_{+,\beta}^{d_{i}}I\tilde{\Phi}^{\mu
\nu} = 0
\end{eqnarray}
Consider the case when electromagnetic or gravitational waves propagate in
the space with $a_{0}\approx constant$, i.e in the domain of  space that
far away from the sources of these fields. If neglect by the members
describing the interactions with the fields with an imaginary masses and
replace the functions $A_{\mu}$ and $\Phi^{\mu\nu}$ by the plane waves
with frequencies $\omega$ we obtain (using (\ref{5}))
\begin{equation}\label{16}
  (1+a_{0})\frac{\partial}{\partial t}F + \frac{\partial a_{0}}{\partial
  t}F - {\bf k}^{2}=0
\end{equation}
\begin{eqnarray}\label{B} \nonumber
F=(1+ a_{0})\frac{\partial}{\partial t}A_{\mu} +
  \frac{\partial a_{0}}{\partial t}A_{\mu}
\end{eqnarray}
or
\begin{equation}\label{17}
\omega^{2} -  2i \frac{\partial a_{0}}{\partial t} \omega -
(\frac{\partial^{2}a_{0}} {\partial t^{2}}+ (\frac{\partial a_{0}}
{\partial t})^{2})- {\bf k}^{2}c^{2}=0,
\end{equation}
For  $\omega$  obtain
\begin{equation}\label{18}
 w= i\gamma  \pm\sqrt{\frac{\partial^{2} a_{0}}{\partial t^{2}} + {\bf k}^{2}c^{2}}
 \quad\quad\gamma = \frac{\partial a_{0}}{\partial t}
\end{equation}
So for $|k|=0 $ or ${\bf k}^{2}c^{2}$ $<< \frac{\partial^{2}
a_{0}}{\partial t^{2}}$ we read
\begin{equation}\label{19}
\omega_{0} = \sqrt{\frac{\partial^{2} a_{0}}{\partial t^{2}}} +i\gamma
\end{equation}
and for $\frac{\partial^{2} a_{0}}{\partial t^{2}}$  $ << {\bf
k}^{2}c^{2}$ obtain
\begin{equation}\label{20}
\omega = |{\bf k}c| + 0.5 \frac{\omega_{0}^{2}}{|{\bf k}c|} +i\gamma
\end{equation}
Thus  it was shown that the massless fields got the rest masses and these
masses defined by the fractal dimensions of the time. If  the fractal
corrections to the fractal dimensions of time are equal zero the rest
masses are also equal zero. The value of rest masses depends of the FD of
time $ (d=1+\varepsilon({\bf r}(t),t))$. So the rest masses are functions
of time and coordinates included by means of the densities of energies
$\beta_{i}L_{i}({\bf r}(t),t))$ and differs for different time and in
different coordinates. This value and its derivatives with respect to $t$
gives the rest masses for case when its velocity is equal zero. In fractal
theory of time and space a masses may have any velocities. Now we must
take into account this possibility and use the relation for moving
particles (see \cite{kob4}) in modified SR for multifractal time
\begin{equation}\label{21}
  E=\beta^{*-1}E_{0}=\frac{E_{0}}{\sqrt[4]{(1-\frac{v^{2}}{c^{2}})^{2}+4a^{2}}}
\end{equation}
where $a^{2}=(\frac{\partial\varepsilon}{\partial t})^{2}t^{2}$. Thus for
the rest  masses  originated by the fractal dimensions of time for the
case when its velocity $v$ equal speed of light $v=c$ obtain
\begin{equation}\label{22}
  \omega_{c}=\frac{\omega_{0}}{\beta^{*}}= \sqrt{\frac{\omega_{0}^{2}}{2a}}
\end{equation}
The considered theory do not define the values of parameters $t$ and
$t_{i}$ included in relations (\ref{25})-(\ref{26}), so any estimations of
the values of rest masses of gravitons and photons are preliminary and
depends of hypothesis laying in the methods of selection $t$ and $t_{i}$.
Nevertheless, as example, consider the order of the rest masses of
gravitons and photons (neutrinos also have the rest masses originated by
FD of time, but it seems that the main part of the rest masses of
neutrinos may be explained by ordinary theories).\\
\\ a) gravitons masses\\
\\ Let $a_{0}= \varepsilon$ and $\varepsilon$ with grate accuracy may be
treated as constant value. Than
\begin{equation}\label{23}
\frac{\partial\varepsilon}{\partial t}\approx \frac{\varepsilon^{2}}{t}
\end{equation}
and for large $t$ ($t$ is the time of existence of Universe)
$\frac{\partial\varepsilon}{\partial t}$ also may be treated as constant
value. In this case
\begin{equation}\label{24}
\frac{\partial^{2}\varepsilon}{\partial t^{2}}\approx
\frac{\varepsilon^{3}}{t^{2}}
\end{equation}
and for the rest mass of gravitons $m_{g}$ we obtain
\begin{equation}\label{25}
    m_{g}\approx \sqrt{\frac{\varepsilon}{tt_{i}}}\hbar c^{-2}
\end{equation}
For $ \varepsilon \sim \frac{r_{0}}{r}$ where $r_{0}$ and $r$ are the
gravitational radius  of Earth and its radius, $ t\sim10^{16}sec$ and
 $t_{i}\sim 10^{-34}sec$ ($t_{i}$ is the time needs for graviton  to
receive the velocity of light) is the time of "Big Bang" origin, we obtain
$ m_{g}\sim10^{-43}g$. If differentiate
$\frac{\partial\varepsilon}{\partial t}$ with respect to $t$ then for
estimation of $m_{g}$ we receive
\begin{equation}\label{26}
 m_{g}\approx \sqrt{\frac{1}{tt_{i}}}\hbar c^{-2}
\end{equation}
and  $m_{g}\sim 10^{-39}g$.\\ \\ b) The rest masses of photons\\ \\ We may
use the same assumptions for estimation  the rest masses of photons that
have been used for  the estimation of the rest masses of gravitons. So we
receive the same formula. The estimation of $t_{i}$ will be the same that
for gravitons ($m_{ph}\sim10^{-39}g$). The  another  selection of $t_{i}$
as the time of emitting of quanta of light $t_{i}\sim10^{-12}sec.$ gives
$m_{ph}\sim 10^{-50}g$. We pay attention, as $\varepsilon$ is in general
not constant value, on not necessity  the using of $\ref{6}$ or $\ref{8}$
for calculation the derivatives of $\varepsilon$ with respect to $t$ if is
known the dependence of $\varepsilon$ at $t$.

\section{The problems of hidden masses or black matter in the fractal Universe}

As well known, the theories of galaxies moving needs for the value of mass
of Universe the value almost in nine times large than the  masses observed
by astronomers. So, where is the deficit of matter hidden? Is it the
masses of interstellar gas substance pressed by gravitational fields, or
is it the black matter of unknown nature? These questions do not have now
unambiguously answer. The existence of the rest masses of photons and
gravitons allows to include these masses in the possible scenario of the
models of hidden and black matter. Really, if the masses of photons and
gravitons have the value of order $10^{-39}g$ it is quite enough for
explaining the necessary  deficit of masses in the  Universe.

\section{Can the Bose-condensates of photons or gravitons
exist in the fractal Universe?}

Is it possible to imagine the flow of photons or gravitons with very small
wave numbers and very low temperature that satisfy the conditions that are
necessary for Bose-condensation? It seems ( if its moving take place in
the very strong gravitational or electromagnetic fields that makes its
velocities ( and kinetic energy) very small) because of existence of the
rest masses such condensation is not contradicts the physical laws. The
really answer now is unknown.

\section{conclusion}

The main results obtained in this paper are:\\ \\1. All physical fields
have rest masses originated by the fractal dimensions of time;\\ 2. The
rest masses originated by the FD of time have similar characteristics ( in
our approach ) for any massless fields ;\\ 3. The rude estimation of
values of rest masses gives satisfactory values for explaining the problem
of black matter and hidden masses;
\\4. The really existence of the rest  masses of  photons and gravitons makes
unnecessary aspiration of the formal introduced masses in theories
\cite{log}, \cite{kob1}- \cite{kob15} to zero ; \\ 5. Some questions
arise: have rest masses of photons and gravitons unknown charges?;  can a
charge photons (or a gravitons) interact with each other? What may be
answers on these questions and may the theory of open systems
\cite{klim1}-\cite{klim2} (quantum variants) give these answers?\\6. The
fractal theory of time and space restore the symmetry respect rest masses
for all physical fields

\end{document}